\documentclass[preprint,5p,times]{elsarticle}

\usepackage{amssymb}
\usepackage{lineno}
\usepackage{hyperref}
\usepackage{xcolor}
\modulolinenumbers[5]

\journal{Arxiv}

\biboptions{authoryear}

\makeatletter
\def\@author#1{\g@addto@macro\elsauthors{\normalsize%
    \def\baselinestretch{1}%
    \upshape\authorsep#1\unskip\textsuperscript{%
      \ifx\@fnmark\@empty\else\unskip\sep\@fnmark\let\sep=,\fi
      \ifx\@corref\@empty\else\unskip\sep\@corref\let\sep=,\fi
      }%
    \def\authorsep{\unskip,\space}%
    \global\let\@fnmark\@empty
    \global\let\@corref\@empty  
    \global\let\sep\@empty}%
    \@eadauthor={#1}
}
\makeatother

\begin{document}

\begin{frontmatter}



\title{A Dynamic Web Service Registry Framework for Mobile Environments}


\author{Rohit Verma\fnref{fn1}\corref{cor1}}
\ead{rohitv@iiti.ac.in}

\author{Abhishek Srivastava\fnref{fn1}}
\ead{asrivastava@iiti.ac.in}

\fntext[fn1]{Computer Science and Engineering, 
		Indian Institute of Technology Indore India}
\cortext[cor1]{Corresponding author}

\begin{abstract}
Advancements in technology have transformed mobile devices from being mere communication widgets to versatile computing devices. Proliferation of these hand held devices has made them
a common means to access and process digital information. Most web based applications are today available in a form that can conveniently be accessed over mobile devices. However, web-services (applications meant for consumption by other applications rather than humans) are not as commonly provided/consumed over mobile devices. Facilitating this and in effect
realizing a service-oriented system over mobile devices has the potential to further enhance the potential of mobile devices.
One of the major challenges in this integration is the lack of an efficient service registry system that caters to issues associated with the dynamic and volatile mobile environments.
Existing service registry technologies designed for traditional systems  fall short of accommodating such issues.
In this paper, we propose a novel approach to manage service registry systems provided `solely' over mobile devices, and thus realising an SOA without the need for high-end computing systems. 
The approach manages a dynamic service registry system in the form of light weight and distributed registries.
We assess the feasibility of our approach by engineering and deploying a working prototype of the proposed registry system over actual mobile devices.
A comparative study of the proposed approach and the traditional UDDI (Universal Description, Discovery, and Integration) registry is also included.
The evaluation of our framework has shown propitious results in terms of battery cost, scalability, hindrance  with native applications. 
\end{abstract}



%
%
%
%
%

\end{frontmatter}


\section{Introduction}

Continued evolution in technology has made computing devices an integral part of one's life.
The most common manifestation of this is the `Mobile Phone'. Modern technology has transformed the mobile phone from a mere communication device to a versatile computing device.
These hand-held devices have enabled us not only to access information, but also to provide information to others on the move. Modern mobile phones, equipped with powerful sensors, have endowed capabilities to provide and create near real-time information. This real-time information is useful for oneself and for others. 
An established approach for sharing and provision of information and creating useful applications in a distributed environment is Service Oriented Architecture (SOA)~\citep{Papazoglou2003}. Realizing SOA over mobile devices has the potential to convert mobile phones owned by common people from mere \textit{information subscribers} to \textit{information providers} and beyond.

The major advantage of this is that it can be used in scenarios where there is little or no preexisting infrastructure. Examples of such scenarios include 
War-front, Post-disaster relief management. In such scenarios, mobile based SOA has the potential to  enable ground teams to provide runtime information to commanding units, help teams at disaster sites to exchange data, analyse damage and examine various statistics using mobile devices.
In such systems of SOA over mobile devices all three elements of the SOA triangle: service providers, service consumers, and service registries are realised over mobile devices.

Web services are the proven way towards implementation of a ``Service Oriented Architecture''. 
Advancement in mobile device technology has motivated researchers to explore the possibilities of effectively hosting web services over mobile devices, and thereby trying to realize service oriented systems in mobile environments.
There has been substantial work towards enabling mobile devices to host web services~\citep{srirama2006mobile}\citep{tergujeff2007mobile}\citep{alshahwan2010providing}. 
An important aspect of service oriented systems, ``service discovery'', however, remains a challenge in mobile environments.
There is literature available on service discovery for distributed environments ~\citep{sivashanmugam2004}\citep{kunal2005}\citep{du2006}, but one catering specifically to mobile environments is still lacking. 
Several challenges specific to hosting web services over mobile devices need to be taken into account in such service discovery mechanisms. These include, but are not limited to battery and network constraints, limited computational power of mobile devices. 
Moreover, such dynamic mobile services are prone to uncertainty (owing to network outage, battery issues, physical damage) and frequent changes in functionality (primarily owing to the change of context), and hence make frequent service updates a necessity to effectively function as web-services.
The role of the \textit{service registry} therefore, becomes one of prominence to properly 
manage such dynamism. 
Traditional service registry solutions for web-services such as UDDI~\citep{uddi302}, ebXML~\citep{hofreiter2002ebxml}, can not be directly utilised in such environments that require frequent updates. 
What contributes to this is the exhaustive data model of such registry offerings that is hard to analyze and parse for mobile devices at run-time. 
To the best of our knowledge, the current work is the first attempt to comprehensively investigate these issues and design a dynamic service registry that facilitates service discoveries in mobile environments.  

As mentioned earlier, the ultimate aim is to realize a service oriented architecture over mobile devices without involving high end servers. 
Hence, the proposed architecture provides all registry related information and operations using mobile devices itself, without requiring high-end computers or high management costs. Further, in order to support scalability, fault tolerance, and fault localization, we propose a distributed and category based service registry. 

To demonstrate the feasibility of the approach, we have engineered a prototype deployment.
 This includes heterogeneous and loosely coupled mobile devices deployed in a collaborative manner to manage the service registry along with native hosted services. 
We also compare the proposed approach with the traditional UDDI system for managing service registry from the perspective of mobile devices. The evaluation shows propitious results in favour of our approach wherein the latter is shown to have acceptable battery requirements, low data communication costs,  promising scalability, and little or no hindrance to the working of native applications of mobile devices.

This work is a significant extension of our previous work \citep{rohit2014ws}. 
In our previous work~\citep{rohit2014ws}, we had introduced an XMPP based model to maintain a service registry for mobile environments. The main focus of the work was to introduce the registry architecture and the communication mechanism followed during service discovery from the registry. In the presented work, we provide a holistic service registry framework that makes use of XMPP based service registry framework at the core. 
We defined the roles for mobile devices involved in the service registry framework to provide a scalable mobile registry solution.
We have further extended the service registry operations to cater the specific needs of the mobile environment. 
We further provide detailed descriptions of the various registry operations that facilitate the realisation of a dynamic mobile service registry. We further evaluated the proposed approach by realizing it through a working prototype and deployed it over mobile devices of volunteers. We further present a detailed literature survey that covers various categories of service registries.

The rest of the paper is organized as follows: Section~\ref{sec:motive} presents a motivation and requirements for the novel approach. Section~\ref{sec:arch} provides details of the proposed approach and design concepts. 
Various registry operations are discussed in Section~\ref{subsec:operation}.
Prototype implementation details and inline comparison with UDDI are presented in Section~\ref{sec:implementation}. Section~\ref{sec:evaluation} includes notes on the experimental evaluation of the approach. This is followed by Section~\ref{sec:related} that presents a survey of related work. Finally, Section~\ref{sec:conclude} concludes the paper with a brief discussion on future possibilities.

\section{Motivation}
\label{sec:motive}

Kotler et al.~\citep{kotler1996} suggested services as ``activities or benefits offered for sale that are essentially intangible and do not result in the ownership of anything''. A mobile service defined in this work is a service that is offered from mobile phones of providers; this may also include information provided by the mobile sensors, third party software, or human users.
This allows different machines to exchange information with each other over a network, without necessarily requiring a user interface. In general, the service may be a component or sub-part of the web application that is usually used by human users. For example, a chatting web application provides GUI to human users to communicate with another human. While a presence service embedded in the web application detects the presence of other machines, this presence service does not require any human intervention.
%

\subsection{Motivating Scenario}
\label{subsec:scene}


\textit{Alice is a high risk cardiovascular patient. Recently, she got an ECG sensor implanted in her body~\citep{ban2014survey} that monitors her cardiovascular health and provides statistics and information as a mobile service via her mobile phone. This service can be consumed by her cardiologist and she can be provided with proper prescriptions as per her current health. One day she had a sudden cardiac arrest on her way to another city. Alarming variations in her ECG signals were observed by the service on her mobile device and the service discovered the nearest ambulance through the latter's exposed mobile service. 
Further, her mobile service automatically provided access to her latest ECG signals to the ambulance support medical staff and enabled them to prepare well in advance for the patient.
The ambulance was able to discover her current location through another service on her mobile device that provided GPS coordinates. 
Further, when the ambulance was on its way, the ambulance's mobile service provided the doctors at the nearest hospital with the latest information on the situation. 
Simultaneously, the hospital was able to  make use of Alice's ECG mobile service to gather her ECG history and prior to her arrival the doctors at the hospital had a chance to study her medical profile and case in detail.
On its way to the hospital, the ambulance was able to make use of the services  exposed by other travelers on their respective mobile devices to avoid the busy route and opt for the path with less traffic.
Meanwhile, the insurance company was contacted by Alice's mobile service and her hospital information was provided, so that the financial aspect could be taken care of even before her arrival. Alice's cardiologist was also able to provide details of his/her prescriptions via his/her mobile service to the doctors in the hospital so that the latter could learn about her medications and allergies if any. 
}

With rapid advancements in mobile technologies and wireless networking, mobile devices have become perhaps the most suitable and economical solutions for the provision of dynamic, transient, contextual, personalized services.
These mobile provisioned services can make the service access handy and convenient for service consumers. Further, the provisioning of mobile services is an economical solution that requires little or no pre-existing infrastructure. 
In the discussed scenario, Alice, her cardiologist, the ambulance, hospital staff can make use of each other's mobile services in critical situations and can provide assistance to Alice. This explains the importance of mobile services;
subsequently, however, the above scenario also raises a question: "How is the mobile service consumer able to discover the appropriate service among such large number of services and that too in an uncertain environment as a mobile environment?". 

\subsection{Need for a Novel Mobile Service Registry}
\label{subsec:need}

Web services hosted on mobile devices are mainly useful for sharing contextual, personal, proximal information. 
Mobile devices in such environments are mostly distributed arbitrarily and make service discovery and management of service registry a cumbersome process. In this section, we 
discuss the need for a novel service registry architecture for mobile environments.

In order to provide an effective service registry for mobile environments, two approaches are possible. The first is a classical centralized service registry approach where all information on the available mobile services is maintained at one place and this is usually over a powerful computing device; The second approach is a decentralized service registry approach. Here, the registries are maintained by a system of distributed nodes in such a way that each node caters to a fraction of the services and there is a large degree of redundancy.  Between these two approaches, the decentralised service registry approach appears to be more   
appropriate for mobile environments. There are several reasons for this such as the issue of a single-point-of-failure in the case of a centralised system, the lack of a definite guarantee of continuous reliable connections between mobile devices and the central server, the difficulties of rapid and regular updates in large centralised registries thus giving rise to  obsolete information and so on. There are, of course, drawbacks in the decentralised system as well. It is these drawbacks that we will discuss and attempt to overcome in the rest of this paper. 
Cloud offloading is another approach that is often used for facilitating services over mobile devices.
In cloud offloading, the service logic, usually, resides on the cloud and the mobile devices may work as the proxy for these services.
The associated concerns of cloud offloading~\citep{Sanaei2014survey} (significant network delay and latency, rigid SLA requirements etc.) however do not make it a potential candidate for dynamic mobile service registry. Sanaei et.al.~\citep{Sanaei2014survey} discuss these challenges in detail.

A decentralised service registry may be realised using either traditional service registry approaches that are commonly used in legacy wired systems (such as UDDI~\citep{uddi302}, ebXML~\citep{hofreiter2002ebxml})  or a new approach especially catering to the vagaries of mobile environments may be adopted.

Though possible, adopting traditional registry approaches such as UDDI~\citep{uddi302} from W3C 
is ill suited to dynamic mobile environments. 
The traditional registry architecture comprises 
UDDI data entities (businessEntity, businessService, bindingTemplate, tModel, publisherAssertion, subscription), various UDDI services and API sets, UDDI Nodes for supporting node API set, UDDI Registries. 
Such a base architecture is quite `heavyweight' and makes it difficult to host UDDI over mobile and resource constrained devices.
Further, services offered over mobile devices tend to behave in an anarchic manner; as the changes in the functional, non-functional, other aspect of the services may be quite frequent owing to regular change in context and networking environment of the device. This requires frequent updates to the service registry.
UDDI, on the other hand, is designed around concepts of SOAP/WSDL, heavyweight technologies that make frequent updates a cumbersome process.
As a consequence, the information on the UDDI registry quickly becomes obsolete.
A new approach, therefore, is imperative for maintaining an effective registry system for mobile environments. 

\subsection{Mobile Service Registry Requirements}
\label{sec:req}

Before we get into detailed discussions on the proposed approach, here is a quick point-wise summary of the requirements for effective service registries for mobile environments. This is along the lines of Dustdar et al.~\citep{dustdar2005view} who did something similar for articulating general requirements for web service registries. 

\textit{R1: Management of transient web services:} 
The very nature of mobile devices makes hosted web services repeatedly and randomly enter and leave the network.
A service registry should be such that it supports such dynamic and frequent  
arrival and departure of service providers. 

\textit{R2: Lightweight:} 
A service registry designed for mobile environments should be lightweight. 
A lightweight service registry would complement the power (i.e. battery) and computational constraints of mobile devices.
Furthermore, a lightweight service registry is agile and is easier to integrate with diversified mobile environments. 

\textit{R3: Minimum communication overhead:} Given the battery and network constraints in mobile devices, emphasis should be towards a registry system with minimum communication overhead. 

\textit{R4: Distributed service registry:} As the number of mobile devices (and therefore potential web services over these)  are increasing exponentially, a centralized service registry system has limited utility and gets outdated very quickly.
Hence, a distributed service registry system is required to support scalability. 

\textit{R5: Enabling run time search:} An important enabler of mobile based service oriented systems is support for run time search. This is necessary owing to the frequent arrival of new and often more competent services and/or failure of existing services.

Conforming to the above points could potentially ensure a service registry suitable for mobile environments.

\section{Proposed Approach}
\label{sec:arch}

In mobile based SOA environments, each mobile device can perform the functionalities of both a service provider and a service consumer. As a service consumer, a mobile device discovers web-services and invokes them after negotiating with the providers. As a service provider, a mobile device hosts services and publishes hosted services with service registries. However, as stated earlier, an effective mobile registry to publish and discover such mobile services in dynamic environments is still lacking.
The mobile services are provided by mobile devices and may be consumed by another mobile device in a peer to peer manner.

\subsection{Registry Details}
\label{subsec:registry}

Our approach suggests a service registry system that comprises a light-weight registry server at each participating mobile device. 
The registry at each mobile device contains minimal information that is \emph{just} sufficient to uniquely identify the registered entity. A registry server (at each mobile device) manages either of two types of registries: 

\begin{enumerate}
\item Service Registry: Registered mobile services are managed in the service registry. The service registry contains an entry for each service as:
\textit{service name, service access point, service ID, service description, service groups, availability, service location, service provider, other service information}.

\item Group Registry: Registered services are categorized into service groups. These service groups are managed in a group registry.  The group registry contains the following information for a service group: \textit{group name, group domain, group description, registrant, groupid, group access point, other group information}. 
\end{enumerate}

\begin{figure}[t]
\centering
\includegraphics[width = 7.5cm, height = 8.5cm]{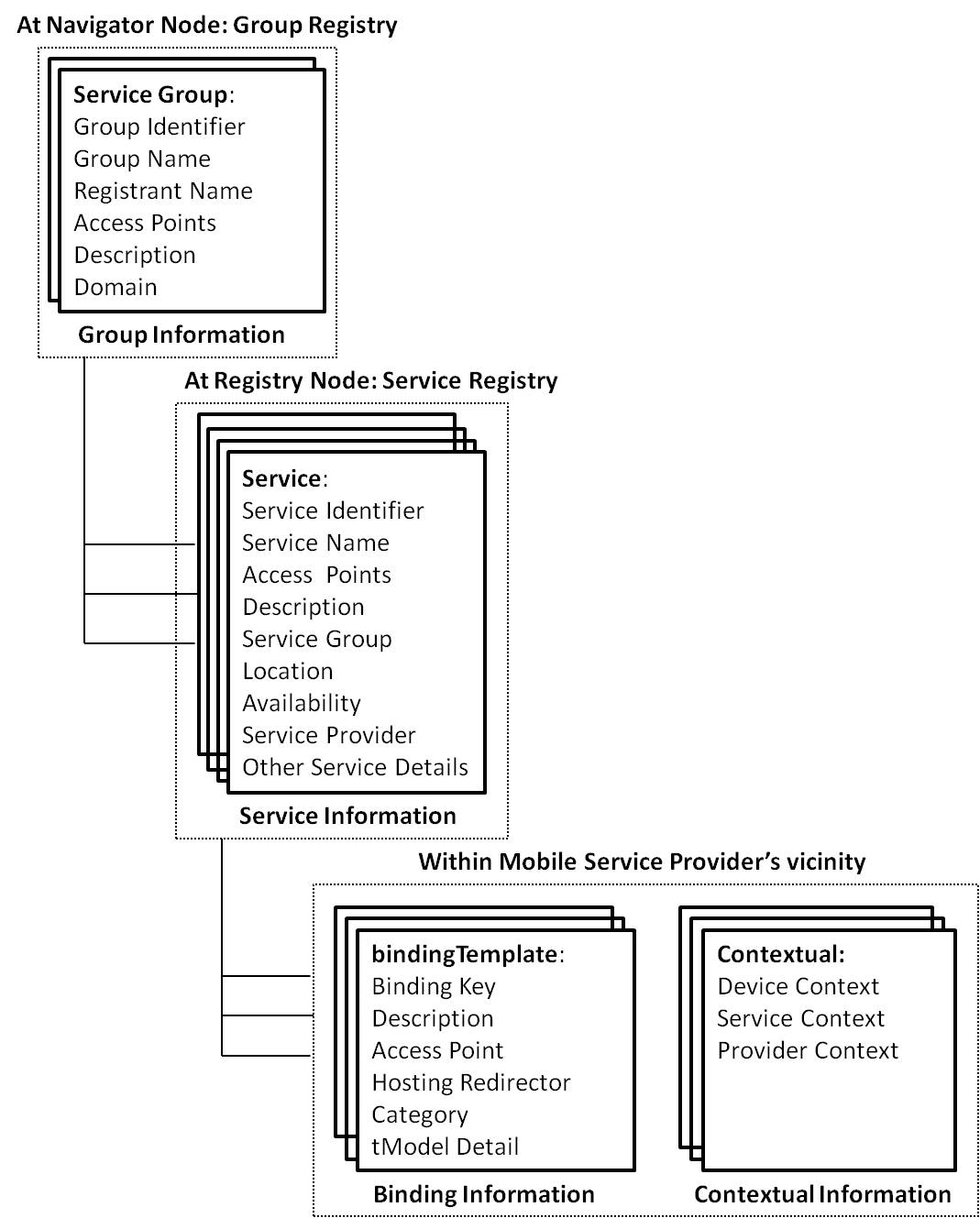} 
\caption{Mobile Registry Entries}
\label{fig:regent}
\end{figure}

The organization of registries, as proposed, is shown in Figure~\ref{fig:regent}. 
The service information that is just enough to identify a registered service and that is less likely to change is kept in the registries
and service information that is more likely to change but does not affect the discovery process of the service is kept in the vicinity of the provider.
The service binding description, and contextual descriptions are provider specific and are likely to change in the mobile environment.
Therefore, these descriptions of the service are kept in close vicinity of the mobile service provider. 
Close vicinity here implies that the description is hosted on the same mobile device as the service or a third party repository, where these descriptions can be updated rapidly. 

We define several registry related operations in Section~\ref{subsec:operation} that are performed using XML streams. 
These XML streams are inspired by XMPP~\citep{saint2004extensible} (eXtensible Messaging and Presence Protocol), a well known and established communication protocol. XMPP is already in wide use in mobile environments in several instant messaging applications. 


The proposed approach provides service registry operations that facilitate effective discovery of a service. 
Details like the non-functional descriptions and quality of service values of the services have deliberately been kept out of the proposed system to make it as 'lightweight' as possible and hence suitable for mobile environments.

\subsection{Design Concept}
\label{subsec:concept}

\begin{figure*}[!htbp]
\centering
\includegraphics[width=18cm, height=10.3cm]{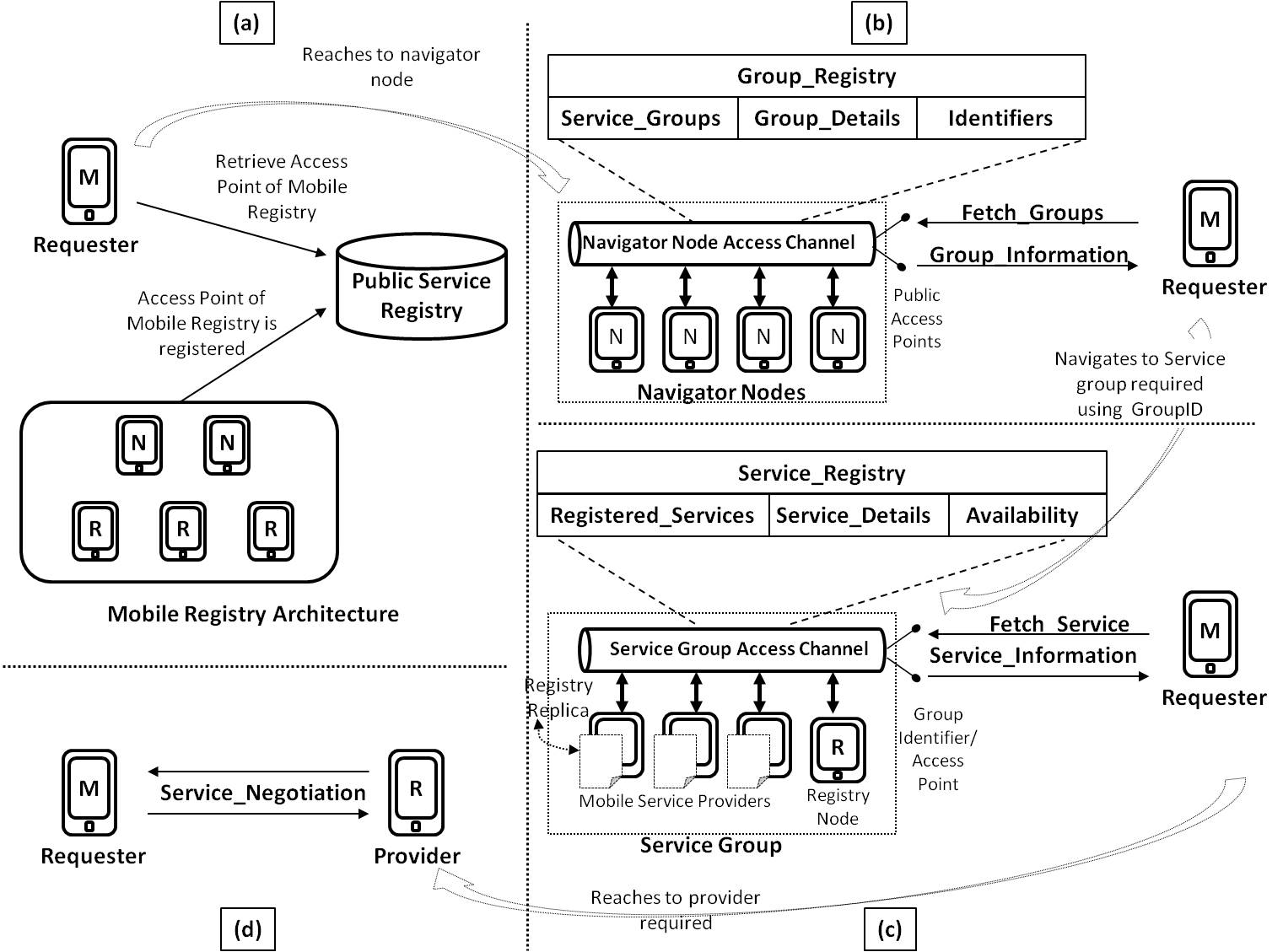} 
\caption{Mobile Service Registry Approach - (a) Retrieval of Mobile Registry Access Point (b) A requester fetches for service group at the Navigator Node (c) A requester fetches service required at the Registry Node (d) Requester contacts directly service provider returned from the Registry Node.}
\label{fig:approach}
\end{figure*}

There are four primitive steps in the proposed mobile service registry approach that are presented in Figure~\ref{fig:approach}: 

1) \textit{Mobile service registry access point is retrieved:} As shown in Figure~\ref{fig:approach}.a, a registry requester (represented by a mobile icon with \textbf{M}) accesses the public registry to retrieve the access point details of mobile service registry. 2) \textit{Mobile service registry is accessed via Navigator Nodes:} As shown in Figure~\ref{fig:approach}.b, the navigator nodes (represented by mobile icons with \textbf{N}) are contacted for the ``Group Registry''. This Group Registry contains the list of service groups. Service group of the required service provider is discovered in the group registry. 3) \textit{Service Group is contacted via Registry Nodes:} As shown in Figure~\ref{fig:approach}.c, the registry nodes (represented by mobile icon with \textbf{R}) are accessed via groupid for retrieving the service provider's information. ``Service Registry'' is traversed for the required service provider. 
4) \textit{The service provider is contacted:} As shown in the Figure 2.d, finally, the required service provider is discovered and it is contacted for service negotiation and service binding.

These steps are discussed in detail in the next subsection. We first start with the roles performed by the mobile devices.
A mobile device potentially performs the following roles (as shown in Figure~\ref{fig:approach}.a): Navigator Node and Registry Node.

\subsubsection{Navigator Nodes}

Navigator nodes are the entry points for the mobile registry architecture as shown in Figure~\ref{fig:approach}. 
These navigator nodes are accessible by service consumers via public access points. 
We have devised the mobile registry architecture as a service itself. 
The mobile registry and its public access points can be registered with any global public registries just to make them globally discoverable (as shown in Figure~\ref{fig:approach}.a).
The motive to use global public registry is to provide the access point details of the mobile registry architecture; mobile devices would need to use the global public registry just once to retrieve the 
access point details of the architecture.
(These global public registries could be any existing service registries as discussed in section~\ref{sec:related}. These registries are assumed to be well in place, hence is not discussed in detail.
The global public registry is not suitable for mobile services, the reasons are already discussed in section~\ref{subsec:need}.)
There can be multiple navigator nodes connected to the public access points via a common access channel, as shown in Figure~\ref{fig:approach}.b. The common access channel can be viewed as a communication bus that enables various mobile devices to communicate. 
The common access channel gives mobile devices the liberty to join and leave the network at any time without disturbing other navigator nodes.
Whenever a new mobile device joins as a navigator node, the group registry is updated/downloaded 
via the access channel and the shared domain ontology becomes accessible. The idea of a common access channel can be realized using existing networking technologies as suggested in RFC1112 and RFC5771.


Navigator nodes are the mobile devices that manage the group registry (refer Figure~\ref{fig:regent}).
As discussed in the earlier section, the group registry manages service groups.
These service groups are uniquely identified
by group identifiers or groupid. 
The group registry comprises the list of service groups present in the network along with the respective group identifiers and other group details. The navigator nodes are further responsible for categorizing the registered services into the various service groups and to navigate the service providers to the assigned service groups.

Navigator nodes rely on existing ontological approaches to categorize services on the basis of their domains of offering (or offered service type). 
All navigator nodes have access to the shared domain ontologies~\citep{sabou2005learning} for categorization of the offered services.
Whenever a service provider needs to register its offered service, the group registry is referred first for matching the service with its service group. 
In case the service offered by the provider does not belong to any of the existing service groups, a new service group is created by referring the domain ontology and updating the group registry with a new group entry. 

We have used an existing classification method for classifying services into groups~\citep{allahyari2014}. 
The motivation behind adopting this method is that the method does not require a training set for classification and dynamic changes to the classification parameters is possible without having to retrain the classifier. This is of particular importance in mobile environments as it provides run time updates to the domain ontology without disturbing the classifiers. 

The service group classification method follows generic steps. First, the mapping criteria are parsed from user supplied service description which is in the textual format. Then, the domain ontology is mapped to the mapping criteria. The process calculates a matching categorization score of the service description with the ontological context as defined by Allahyari et al.~\citep{allahyari2014}.


For example, the mobile service providers hosting services for: a.) Doctor's rating and b.) Hospital building floor map, share the same service group \textit{Hospital}, hence they are identified by the same \textit{groupid}. 
However a new mobile service provider offering contact information of pizza outlets 
would fall in a separate service group. We do not dwell upon the classification approach in this paper. The interested reader is referred to~\citep{allahyari2014} for more details on this.


\subsubsection{Registry Nodes}

Registry nodes are the mobile devices that manage the service registries (refer Figure~\ref{fig:regent}). 
As discussed in the previous section, a service registry manages the registered services that are uniquely identified by service identifiers. This enables a service provider to provide multiple services over a single mobile device.
The service registry comprises the list of registered services in a service group, their availability information along with the service details that are just sufficient to manage and identify the registered services. 
Of these details, the real time availability information of the registered services is what mainly contributes to overcoming the uncertainty of mobile environments. This availability information managed at the registry node gives the much needed reliability to the services hosted on mobile devices.
Registry nodes are responsible for managing the up-to-date service registry, responding to service registry related queries, and performing registry related operations.


The service group can be seen as an overlay group of registered mobile service providers and registry nodes that are identified by the groupid. 
We have devised this group identifier as a multi-cast address for the service group members.
The requests sent to the service group are received by all the member mobile service providers, however, only the group member acting as the registry node responds to the requests (as shown in Figure~\ref{fig:approach}.c). 




To improve query response time, a replica of the service registry that is retrieved from the registry node is managed at all the mobile devices hosting services in the service group. 
Selective updates are performed to keep the local replica updated.
During service discovery, the local replica is first referred to, in case discovery fails at the local replica then the registry node is contacted. 
Maintaining replicas of this kind do add a little overhead to the architecture but in the larger context the local
replicas reduce traffic meant for discovery over the network substantially.  Further, the local replicas ease the transition of service providers into full fledged
registry nodes in the eventuality of a registry node failure (details on this in the subsequent subsection).

\subsubsection*{Failure Management:}

Mobile devices acting as navigator nodes or registry nodes can also depict uncertain behavior and are prone to failure.
Hence, in the case of existing registry node failure, a new registry node can be elected from the member mobile service providers, without compromising on the consistency of other service groups or navigators. 
(The same approach is also applied to navigator nodes.) 
Heartbeat operations are used to detect registry node failure.
Any mobile service provider can become a registry node by participating in an election and declaring its candidature. 
Our method of registry node election is inspired by the leader election problem of distributed computing. The algorithm is  discussed in more detail in Section~\ref{subsec:mobops}. 



\subsubsection*{Use-Case Scenario:}
\textit{Potential service consumers} use the proposed mobile registry architecture for discovering their desired services.
This service discovery is a three step process. 
In the first step as shown in Figure~\ref{fig:approach}.b, the service consumer sends the discovery request to the mobile registry architecture via a well known access point or URI (Uniform Resource Identifier) for the desired service. 
This request for service discovery is first handled by the navigator node.
The Navigator node searches its group registry and provides the matching service group details along with the groupid for the requested service. 
This \textit{groupid} acts as a multi-casting address for all the service providers that belong to the group.
In the second step as shown in Figure~\ref{fig:approach}.c, the service consumer contacts the service group using the groupid of the required service. The Registry node of the service group responds with the available matching services and their corresponding service details. 
In the third step as shown in Figure~\ref{fig:approach}.d, service consumer contacts the mobile service provider offering the required service to retrieve the technical description and performs the service negotiation for service access.

\textit{Mobile service providers} use the proposed mobile registry architecture for registering their offered services. The service registration process primarily comprises two steps. First, the mobile service provider sends a service registration request to the mobile registry architecture via the well known access points. The Navigator nodes first handle the registration request and respond with the matching service group along with the groupid and other group details registered in the group registry.
Second, the mobile service provider contacts the service group and registers its service with the service registry of the registry node. Alternatively, if there is no matching service group in the group registry, the navigator node creates a new service group. In this case, the registrant mobile service provider becomes the registry node of the newly formed service group.

\section{Mobile Registry Operations}
\label{subsec:operation}
In this section, we describe the operations and functionalities that provide registry operations such as registration, discovery, service updates, and service binding in the proposed architecture. An inline comparison with UDDI is also presented. The following registry operations make use of several components discussed in an earlier work of ours~\citep{rohit2014ws}.

\subsection{Basic Registry Operations}

\subsubsection{Registration:}
\label{subsubsec:reg}

\begin{figure}[h]
\centering
\includegraphics[height = 6.4cm, width = 8.5cm]{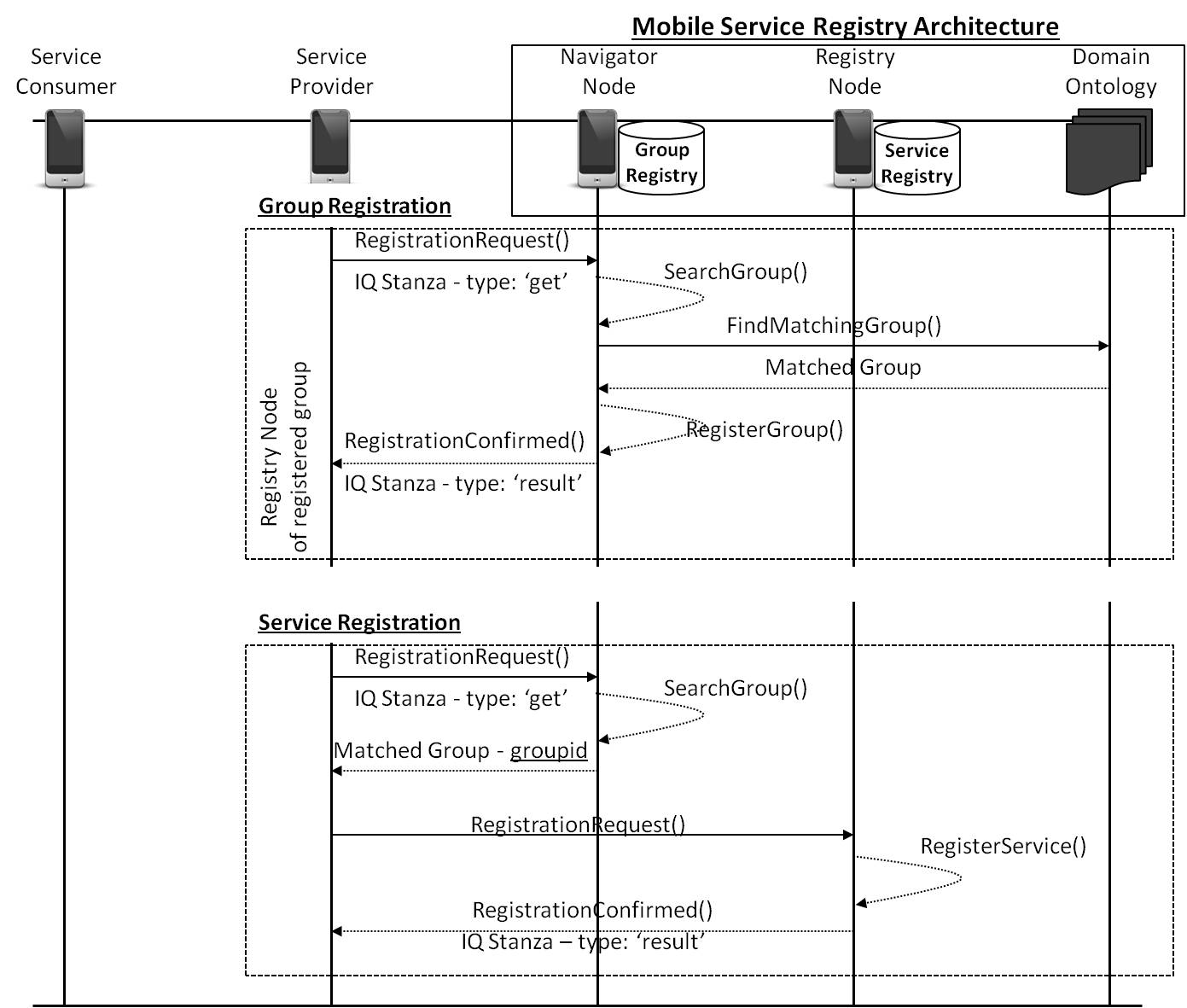} 
\caption{Group and Service Registration}
\label{fig:registerproc}
\end{figure}

In the proposed approach, we have two types of registrations: 1) Group registration (at navigator node) 2) Service registration (at the registry node). These registrations are shown in the Figure~\ref{fig:registerproc}.

\textit{Group Registration :}
The service group is registered in the group registry at the navigator node. 
A mobile service provider (registry client) contacts the mobile registry framework via 
the navigator node to register the service. However, if a matching service group is not yet registered in the group registry, a new service group registration request is initiated.
The mobile service provider sends a group registration request to the navigator node along with the service details.
Hereafter, the navigator refers the domain ontology and based on the service details, a matching group is mapped. This matched service group is updated in the group registry along with its groupid. 
The registry is then shared among all the navigator nodes.
Subsequent to the successful registration process, a ``groupid'' is sent to the newly registered mobile device (in a `result' type IQ stanza). 
At this point the registrant is the registry node of the newly formed service group. Figure~\ref{fig:registerproc} shows the group registration process. 

\textit{Service Registration :} 
Services are registered in the service registry at the registry node. 
The mobile service provider fetches the matching service group at the navigator node and contacts the registry node of the matching service group via the groupid.
Hereafter, the registry node receives service related information from the mobile service provider and generates a serviceid for the new service.
The registration process is completed when the information of the new service is updated at the service registry of the registry node. 
This updated registry is made available to the service providers in the service group for provider initiated updates (pull based updates). Upon successful registration a ``serviceid'' is sent to the newly registered mobile service provider (in a `result' type IQ stanza). Figure~\ref{fig:registerproc} shows the service registration process. \\


%

\noindent\textit{Web Service Registration in UDDI:}

Here we quickly discuss the registration process in the traditional UDDI registry system so that one can appreciate the significance of the proposed approach. In UDDI, registration is done using the \textit{publisher APIs set} exposed by the UDDI, such as save\_service, save\_business, save\_binding, save\_t\_model. These APIs are used to save detailed information on the web service, which may not be necessary in case of the mobile based web services. Moreover this information would tend to become heavy for mobile devices to process or transport.

A typical UDDI registry~\citep{alonso2004web} primarily consists of the following information: businessEntity, businessService, bindingTemplate and tModels for a registered service. 
The information is passed on by the web-service provider to the
UDDI registry through publisher APIs (The information is transported as XML tags, we are not showing the XML for the UDDI structure owing to space constraints here. For the benefit of interested readers, we have uploaded details on this at: http://goo.gl/cn8VaP ). Deploying such a UDDI registry over a mobile device would tend to become heavy owing to the limited computational power and network constraints in mobile devices. The UDDI registry would also significantly lag behind in managing the dynamic nature of mobile devices.

\subsubsection{Service Discovery:}
\label{subsubsec:disc}

\begin{figure}[!h]
\centering
\includegraphics[height = 7.5cm, width = 8.5cm]{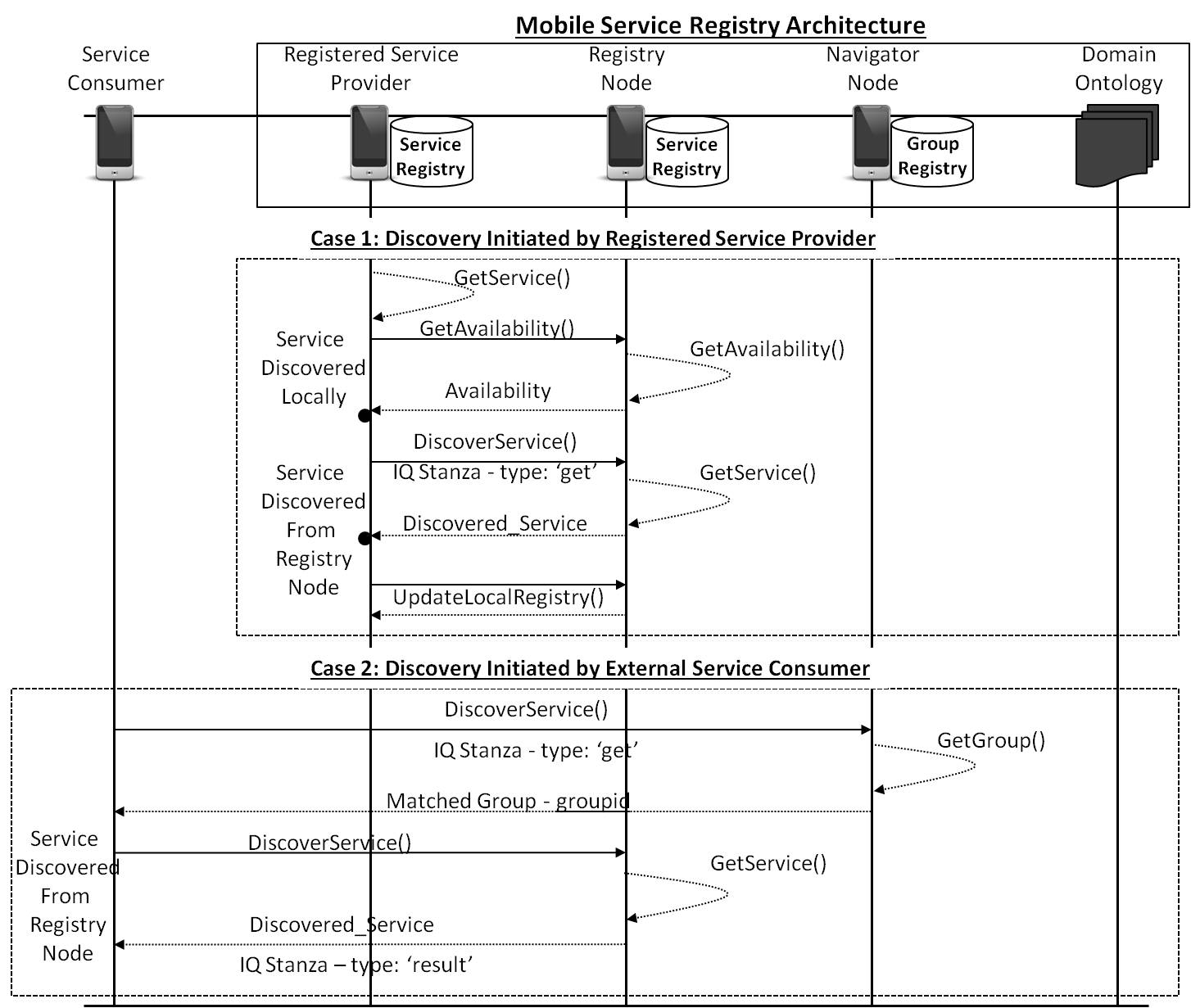} 
\caption{Service Discovery}
\label{fig:registerdisc}
\end{figure}

Two possible cases are included in the prototype: 1. Discovery initiated by the registered service provider, 2. Discovery initiated by an external mobile service consumer. 

\textit{Discovery by registered service provider:} The registered service provider first matches the service group of the required service with the local replica of the service registry. If the group matches then it fetches the required service locally. In case the service is found locally, a selective update is performed from the registry node to obtain information on the latest availability status of the service. In case the service is not found locally, the query is propagated to the registry node and subsequently the local replica is updated with the latest service registry status. 

\textit{Discovery by external service consumer:} The service consumer first contacts the navigator node and retrieves the matching service group information. Subsequent to this the service discovery is forwarded to the matching service group. Afterwards, the registry node of the service group responds with the matching service information. The registered service provider from the other service group also follows this process. Figure~\ref{fig:registerdisc} shows the service discovery process.

\noindent\textit{Service Discovery in UDDI:}

Web service discovery in a UDDI registry is done via
public inquiry APIs of the UDDI, such as find\_service, find\_binding, find\_business, find\_tModel. The service discovery is performed centrally by the UDDI registry server, which requires high computational capability. This is because the consumer 
requests the UDDI registry server which in turn does the query search centrally and responds to the consumer with the results. The complexity and structured nature of the UDDI data structure would makes searching tedious were it adopted in a mobile environment. 
Though traditional UDDIs enable consumers to query the registry and are effective in a centralized system, they are ill suited to the mobile environments that are mostly distributed.

\subsubsection{Service Binding:}
\label{subsubsec:bind}

\begin{figure}[!h]
\centering
\includegraphics[height = 3.5cm, width = 8.5cm]{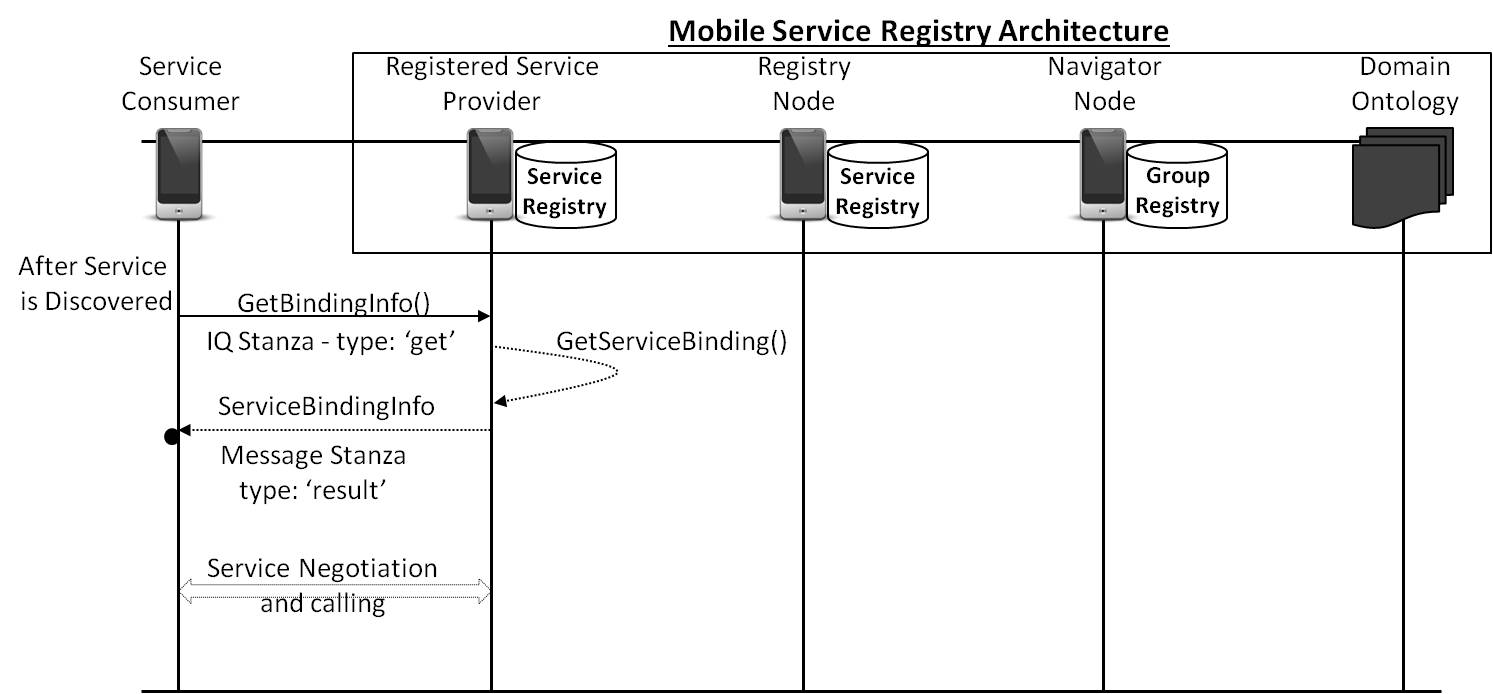} 
\caption{Service Binding}
\label{fig:servicebind}
\end{figure}

Web service binding information is necessary to call a particular web service. It includes the technical information on a web service, such as the access endpoint, required parameter values, return type etc. In the proposed architecture, binding information is exchanged directly between the service consumer and the service provider (as shown in the Figure~\ref{fig:approach}.d). 
The service provider can provide the WSDL/WADL document or it's global URL in the binding information as well. The functional description of the service is kept in the close vicinity of the
service provider. The reason being that the mobile services tend to change frequently and this might result in a change in the technical description of the same. 
Therefore, a proximal location of the functional description facilitates mobile service providers to readily change the service operations dynamically without violating the service registry information (Figure~\ref{fig:servicebind} shows the service binding process).
Furthermore, other types of descriptions viz. non-functional, contextual, business descriptions are usually present on the same device as the service provider to keep descriptions up-to-date without increasing traffic over the mobile registry architecture.

%


\begin{table}[h]
\centering
\caption{Summary of Operations Performed in Proposed Approach and UDDI.}
\label{table:compare}
\begin{tabular}{| p{2cm} | p{2cm} | p{3.6cm} |}
\hline
Operations & Our Approach & UDDI \\ \hline
Service Registration & IQ Stanza & APIs: save\_service, 
save\_business, save\_binding, save\_t\_model \\ \hline
Service Discovery & IQ Stanza & APIs: find\_service, 
find\_binding, find\_business, find\_tModel \\ \hline
Service Binding & Message Stanza & t\_model and WSDL documents \\ \hline
\end{tabular}
\end{table}

\noindent\textit{Service Binding in UDDI :}


Service consumers can retrieve the service binding information of the registered service providers from the UDDI registry using t\_model and WSDL documents.
The t\_model is an exhaustive technical description of the service binding. 
However, due to its inherent complexity service providers often do not update the binding information. 
In fact, some services do not even register themselves owing to this complexity. This has ultimately translated  to
the unavailability of a working and updated global UDDI based registry. 
Nowadays, as a general practice service consumers use web search engines to fetch the binding information. This is done by querying search engines for \textit{filetype} as \textit{wsdl} (for SOAP based web services) or \textit{wadl} (for REST based web services). The results of this searching mechanism leads to all sorts of bias arising out of the indexing and page ranking algorithms of the 
web search engine. Furthermore, the selection would require human intelligence and \mbox{analysis}.

Table~\ref{table:compare} summarizes the discussed operations performed with the proposed architecture and with  UDDI. The operations discussed in the following sub-section viz. presence notification, registry sharing, and registry update are specific to our approach and ones in which no equivalent UDDI operations exist.


\subsection{Mobile Specific Registry Operations}
\label{subsec:mobops}
\subsubsection{Presence Notification:}
\label{subsubsec:presence}


This is one of the novel features of the proposed architecture. 
The presence information of a mobile service is of utmost importance for providing any sort of certainty in 
the volatile mobile environment.
The presence notification provides the current availability 
information on the mobile service. This helps in avoiding as much as possible the failed access of offline services.
Here the terms presence information and availability information are used interchangeably.

The service registry manages presence information for each service.
This helps to uniquely manage the presence information of multiple offered services of a service provider.
The presence information is dynamically updatable and provides availability of a service at a particular instance of time.
The presence information is similar to the availability information in an instant messaging application. 
We managed the presence information as `Available' or `Unavailable' for a service in the presence tag (please refer Figure~\ref{fig:usedxml}). However, this presence tag is a placeholder that can further be 
advanced to incorporate other presence related information.

The proposed approach also incorporates event triggered presence notification that is
generated on the occurrences of events that cause the service to become unavailable. For example low battery level, dropping network strength, critical overload at provider.
These events can be easily detected programmatically using API's exposed by modern operating systems of mobile devices.


\subsubsection{Registry Sharing:}
\label{subsubsec:share}

Registry sharing is required to : 
\begin{enumerate}
\item Manage the latest information about services in registries of various registry nodes. 
\item Keep a local service registry replica at the service providers in service group.
\end{enumerate}





Registry sharing facilitates sharing the registry system over several nodes to form a distributed service registry structure.
Such a distributed registry system becomes particularly useful when a new mobile service provider joins the service group. 
Registry sharing enables the newly joined mobile service provider to retrieve the registry from the group and 
manage a local replica.
Further, the joining of a new mobile service in the service group triggers an update in the registry. 
That needs to be shared with the members of the group in a distributed structure. 
This functionality is extensible and can be adopted for timely updates or event driven updates to manage synchronization in the service registry of the various registry node.

\subsubsection{Registry Update:}
\label{subsubsec:update}

As our approach is mobile service provider-centric and the mobile environment is dynamic, hence a service provider tends to change its configuration on the run.
The registry is required to be updated whenever information on a service gets changed, such as location, access point, service descriptions. 

\textbf{Unregister}: Unregister is a registry update performed when a
mobile service provider discontinues its service offering. Whenever a service provider does not want to provide a hosted service, it unregisters itself from the service registry using the unregister action. 

\subsubsection{Heartbeat Operation:}
\label{subsubsec:heart}

Heartbeats are used to probe various nodes in the proposed registry architecture and to keep the service registry up-to-date. 
A registry node periodically probes other service providers in the service group and
retrieves their latest availability information to keep the service registry up-to-date. Similarly, navigator nodes probe the service group to know if the group is alive and accordingly update the group registry.
Further, heartbeat operations are used for registry node election. New registry nodes can notify the 
service group members about their presence. 

%

\subsubsection{Dynamic Registry Node Election:}
\label{subsubsec:election}

Dynamic registry node election is performed when a registered service provider tries to get promoted to a registry node or when the registry node fails. 
A registry node election is called, when periodic heartbeat signals are not received by the registered providers of the service group.

The registry node election problem is analogous to the leader election problem of distributed computing. Several solutions have been proposed for leader election in distributed computing.
We have adopted an approach for the registry node election that is inspired from Luby's Algorithm~\citep{lubyorigin1986}; a similar approach is described in~\citep{MP2Pluby2008}. The election approach can be implemented in a distributed manner without involving a central authority and produces less message traffic.
In the proposed framework, each service provider announces itself as a candidate for becoming a registry node by 
sending its device details that include battery information, network details, device hardware details, uptime etc. to the existing registry nodes. The service provider with the maximum capability in terms of the mentioned paramenters is selected as the new registry nodes. 
This new registry node sends a heartbeat signal to other registry nodes to make them aware about its new role.

\section{Implementation}
\label{sec:implementation}




We developed a prototype web service implementation based on the proposed approach for mobile devices, using android SDK (ADT 23.0.2) and Oracle Java (version 1.7.0\textunderscore 72). The prototype was built to realize the architecture as presented in section~\ref{sec:arch}. It is worthy to note that \textit{the deployment of the prototype neither required any modification to the device nor did it require root permissions to run}. The prototype developed is independent of the native applications of the mobile device and hosted the web services.
Our approach is applicable to all mobile operating systems, however we chose android for our prototype implementation as it is open source and commands much wider community support.

Our experimental setup comprised seven mobile devices (including Samsung Galaxy S Duos with Android 4.3, Sony Xperia M with Android 4.3, Google Nexus 7 with Android 4.4, Motorola G2 with Android 5.0, three Asus Zenfone 5 with Android 4.4), one laptop (Intel i3 2.13 GHz with 3GB of RAM) and a few running instances of the prototype running on virtual instances of Android devices running on the laptop. These devices comprised the engineered prototype. The setup also had multiple services and service groups. Two experimental wireless networks were setup for the validation. All our experiments were carried for varied network sizes with multiple service providers joining and leaving the network.

\begin{figure}[!h]
\centering
\includegraphics[height = 4.4cm, width = 8.5cm]{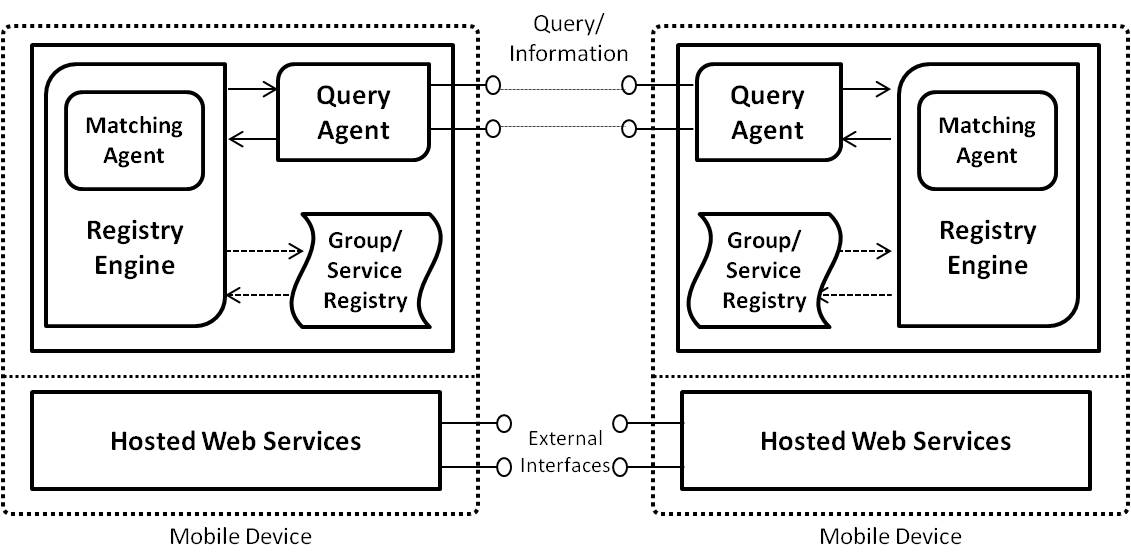} 
\caption{Architecture of Registry and Navigator Nodes}
\label{fig:imparch}
\end{figure}

\textit{Prototye Design:}
The prototype performs all the roles as mentioned in the Section~\ref{sec:arch} - Service Navigator Node, Service Registry Node, and service provider. We have devised registry consumers external and also embedded them with service provider. We have designed a parser to parse and generate the XML streams (Figure~\ref{fig:usedxml} presents the used XML infoset). Figure~\ref{fig:imparch} presents the architecture used for prototype(as discussed in detail in our previous work~\citep{rohit2014ws}).

The presented architecture is deployed on both the nodes: Service Navigator Node, Service Registry Node. The various parts of this architecture are:
1.	Query Agent: This component accepts, generates, and processes the queries from/for other mobile devices. This has external interfaces that can be contacted by any other mobile device. It can be viewed as the XMPP parser for registry management. 
2.	Registry Engine: All the mobile registry related operations as explained in the Section~\ref{sec:arch} are handled by this component. 
3.	Matching Agent: This component parses the result of the service query and evaluates them against the required parameter. Further, it has access to the ontology for service group formation. 
4.	Group/Service Registry: This is the local replica of the service registry or group registry depending on the nature of the node (whether registry node or navigator node). 
The presented approach works even in presence of the hosted mobile web service.

\subsection*{Prototype Specification}
\label{subsec:spec}

This subsection presents identifier and communication related specifications of the prototype for the architecture presented in Section~\ref{sec:arch}. \\

\noindent\emph{Identifiers}: 

Each hosted service over a mobile device was addressed by an identifier. 
This enabled co-existence of multiple services over a single mobile service provider. 
Further, this gave flexibility to the service providers to remove a service without disturbing other services.
We used two types of primitive identifiers: group identifier and service identifier.
An identifier, as discussed earlier, is similar to an email address and is uniquely addressable. The structure of identifier is inspired from XMPP. The format of an identifier is: 

\begin{verb} [GroupID]@[NetID]/[ServiceID] \end{verb}

\begin{verb} NetID \end{verb} specifies the domain or network of a service registry, as there can be multiple service registries that are collocated globally. For private registries, \textit{local} is used as the NetID.
\begin{verb} GroupID \end{verb} is the identifier of the service group. 
\begin{verb} ServiceID \end{verb} identifies the service offered by the mobile service provider. Therefore, there could be multiple unique ServiceIDs for a mobile device offering multiple unique services. 


A service group could be ``TrafficInfo" where the offered services are related to traffic information. A service registry could be for the network: Acme City. This depicts the service group identifier as: \begin{verb} trafficinfo@acmeCity \end{verb}. This service group is shared by all the service providers of service group trafficinfo. A unique identifier for a mobile service provider offering service related to the ``Traffic Information of Main street" could be \begin{verb} trafficinfo@acmeCity/mainstreet \end{verb}.

The identifiers are managed distinctly in a domain by the registry engines and query agents of the mobile devices (as discussed in our earlier work~\citep{rohit2014ws}). Further naming conflicts are resolved by incorporating the service provider`s participation.
In the move to enable networking and realize the proposed architecture over a real network, the identifiers are mapped to the physical network 
using several well accepted networking concepts and technologies. 
These technologies/concepts are available in ad-hoc networking technology, such as \textit{multi-cast domain name system} suggested in RFC 6762, dynamic host auto configuration IP range suggested in RFC 5735 and 3927, multicast networking methods suggested in RFC 1112 and RFC 5771. The libraries for this implementation are  available for the Android OS.\\


\noindent\emph{Communication}:
In the move to provide interoperability over heterogeneous mobile devices, we make use of XML streams for the service registry related communications.
The mobile devices 
communicate and send queries/information in the form of XML stanzas.
These XML stanzas are inspired by~\citep{saint2004extensible,saint1rfc}. 
An XML stanza is the basic unit of communication in XMPP. This discrete semantic unit of structured information or XML stanza is sent from one mobile device to another over an XML stream.

There are mainly three stanza types used in the architecture: \begin{verb} <message  /\textgreater \end{verb}, \begin{verb} <presence  /\textgreater \end{verb}, \begin{verb} <iq  /\textgreater \end{verb}. 
A stanza is a first-level element (at depth=1 of the stream) whose element name is ``message'', ``presence'', or ``iq'' 
The three stanzas are briefly described below. For more details on the stanzas, the interested reader is pointed to our earlier work~\citep{rohit2014ws}.

\begin{figure}[!h]
\centering
\includegraphics[height = 6.4cm, width = 8.5cm]{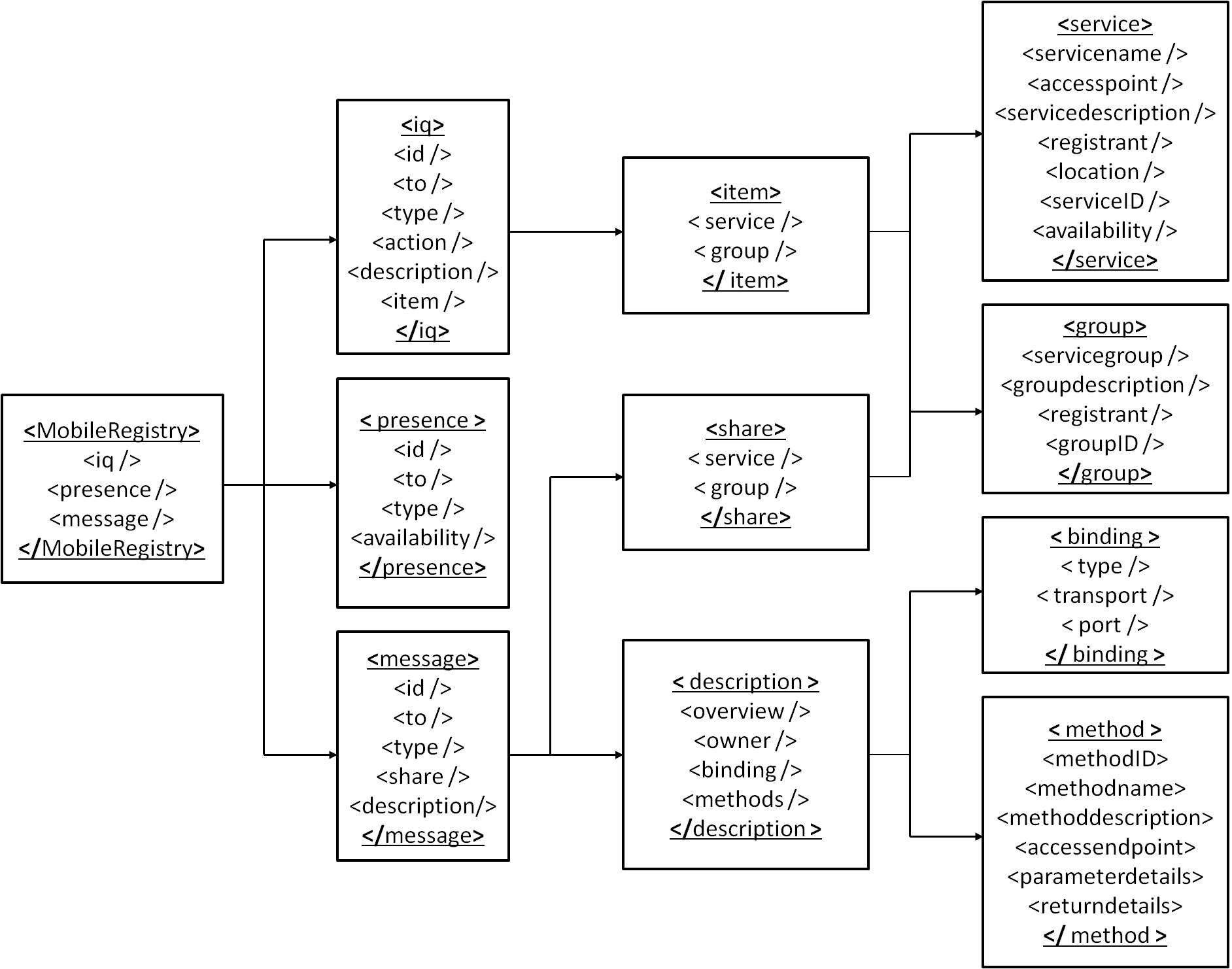} 
\caption{XML Streams Used in Proposed Approach}
\label{fig:usedxml}
\end{figure}

\emph{Message stanza} is primarily used for storing, editing, and sharing service/group information at the service/group registry.
The main purpose that this is managing and updating the service/group registries. 
The Message stanza works in two ways: ``push'' and ``pull''. In push, the information exchange is initiated by the sender e.g. sharing the registry at the registry node with the mobile service providers. Whereas in pull, the receiver initiates the information exchange e.g. selective update of the registry at the mobile service providers. Further, the message stanza is primarily used for service binding operations, registry sharing, registry node elections.

\emph{Presence stanza} is used to update the availability information of services hosted on mobile devices. Each presence stanza includes a brief description and service identifier of the hosted service, along with its availability information. We use `available' and `unavailable' as the primitive presence types in the proposed approach. The presence status of a web service hosted on a mobile device is reflected on the service registry. Presence notification, heartbeat operations are implemented using this stanza.

\emph{IQ stanza} is short for Information/Query stanza. It is based on a request-response mechanism and guarantees a response to a query. The nature of request in the IQ stanza is represented by \begin{verb} type \end{verb}.  Registry information request is represented by \begin{verb} get \end{verb} and it is similar to the HTTP \begin{verb} GET \end{verb} method. 
Any communication involving queries from other mobile devices primarily makes use of the IQ stanza. 
These play an important role in getting information from other mobile devices that host the service registry i.e. registry nodes. 

These stanzas are uniquely identified by the ``id'' element. The ``type'' element represent the type of registry operation that is performed by the mobile devices.
Whereas ``to'' element contains the access point of the recipient (individual mobile device or service group). Figure~\ref{fig:usedxml} shows the detailed XML structure used in the architecture. 
Group and service registration operations, discovery operations, registry update operations are implemented primarily using IQ stanzas.

\section{Evaluation}
\label{sec:evaluation}
\begin{figure}[!h]
\centering
\includegraphics[scale=0.42]{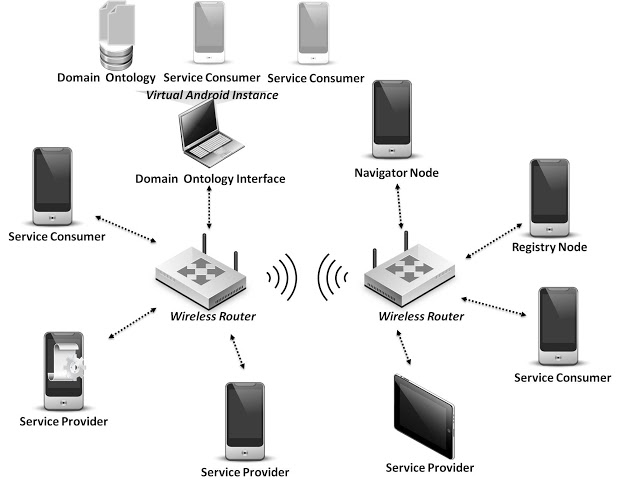} 
\caption{Experimental Setup}
\label{fig:setup}
\end{figure}

To evaluate our approach, we deployed the architecture (presented in Figure~\ref{fig:imparch}) over real mobile devices. 
We solicited volunteer participation to host our prototype over their personal mobile devices. This enabled us to analyze the feasibility of our approach in a practical scenario. 
We established two experimental wireless networks within our institute building to connect the volunteers' mobile devices, laptop, and virtual instances (refer Figure~\ref{fig:setup}).
During the experiment, volunteers were doing their routine work and hence the mobility of the devices followed random patterns. 
We repeated this experiment with varying numbers of
service providers, service registration requests, service discovery queries, with the intent of emulating uncertain situations in practical scenarios. 
In this section, we present the results of the experiment with the motive to show the feasibility of the proposed approach. 

The first experiment evaluated the effect that hosting a registry server had on the battery of the mobile devices. 
We analyzed the effect of various registry operations on the battery. For this purpose, we made use of the power model and solutions suggested by Zhang et al.~\citep{powertutor}. This power model considers the power consumption by a mobile application and also takes into account the application's effect on Wi-Fi, CPU, Cellular interface etc. Table~\ref{table:power} shows the initial battery power consumption by the registry operations. We sent 50 requests of service/group registration, service/group discovery each on the navigator node and the registry node. We observed the power consumption at the navigator nodes and registry nodes by these requests. The values presented in the table are the average battery power consumption. For a quick reference, gmail android app had 567mW, facebook android app had 1297.5 mW, GSM call had 511 mW, and Airplane Mode had 6.4 mW power consumption (depends on build/model of mobile phone). The idea of including these is to emphasize upon the point that the implemention of the proposed mobile service registry solution seems to have acceptable power consumption.

\begin{table}[h]
\centering
\caption{Power Consumption by Various Registry Operations}
\label{table:power}
\begin{tabular}{ p{4.8cm}   p{2.8cm} }
\hline
Registry Operation  &  Power Consumption \\ \hline
Service Registration (For 50 requests) & 44 mW\\ 
Service Discovery (For 50 requests) & 63 mW\\ 
Group Registration (For 50 requests) & 120 mW \\ 
Group Discovery (For 50 requests) & 52 mW \\ 
Heartbeat & 73 mW\\
\hline
\end{tabular}
\end{table}

In our second experiment, we further observed the data bytes being exchanged during these requests. Table~\ref{table:network} shows the total number of bytes exchanged for various registry operations. We concluded that less than a 1KB of data is exchanged for service/group registration. For service/group discovery the number of data bytes exchanged depends on the number of matching groups or services. However, for the purpose of the experiment we had a single matching group or service. That limited the data exchange for discovery requests to under 1 KB for each matching service/group. (Size of an avergae compressed image sent over a messaging app like WhatsApp is 20-25 KB, and 2.52 MB was the background data usage for the Instagram app (with a few account following)  for a day. Therefore, the proposed approach has acceptable data exchange.)

\begin{table}[h]
\centering
\caption{Data Exchanged by Various Registry Operations}
\label{table:network}
\begin{tabular}{ p{4.8cm}   p{2.8cm} }
\hline
Registry Operation  &  Total Data (Received + Transmitted) \\ \hline
Service Registration (For 50 requests) & 48510 bytes\\ 
Service Discovery (For 50 requests) & 41231 bytes\\ 
Group Registration (For 50 requests) & 42161 bytes\\ 
Group Discovery (For 50 requests) & 31487 bytes\\ 
\hline
\end{tabular}
\end{table}

\begin{figure}[!h]
\centering
\includegraphics[scale=0.5]{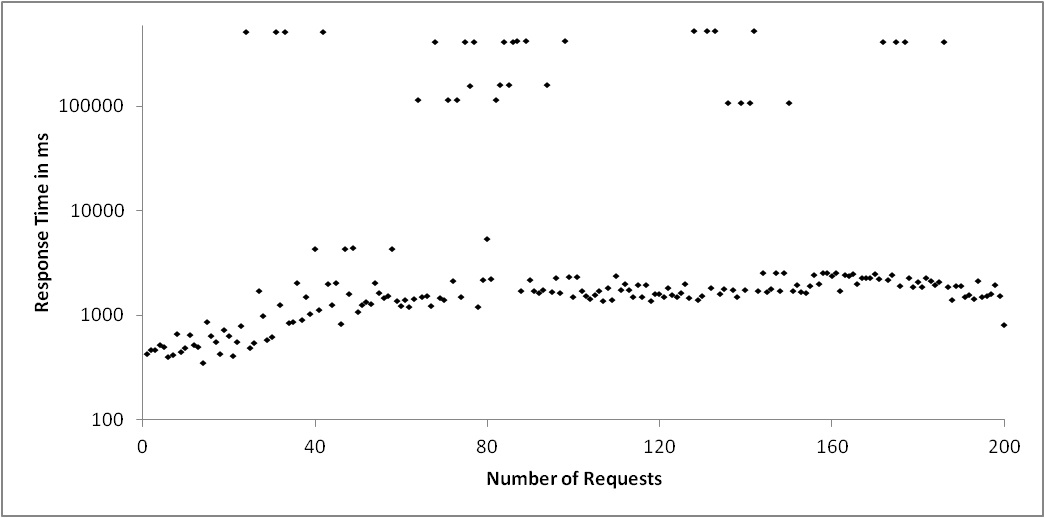} 
\caption{Total time taken for new service registrations}
\label{fig:total}
\end{figure}

In our third experiment, we sent (50*4=) 200 new service registration requests from other mobile devices and virtual instances to the registry architecture. 
Figure~\ref{fig:total} depicts the total response time behavior for a service provider.
Also noteworthy here is the fact that the mobile devices were continuously moving with the volunteers, hence the devices were randomly joining and leaving the network. Therefore, we observed a few outliers in the response time behavior. We concluded that the average service registration time (including outliers) is near 5 seconds which seems acceptable for practical purposes.


The fourth experiment evaluated the effect of directory size on the discovery time. We registered multiple services in a service group on the registry node. 
In order to test the scalability of our prototype, we ran service discovery operations for various directory sizes: 500, 1000, 5000, 10000, 50000, 100000. We discovered that the discovery time increases as the size of the directory increases. However, in spite of this even with 100000 registered services the query response time was under 1 second which is acceptable for all practical purposes.

\begin{figure}[!h]
\centering
\includegraphics[scale=0.5]{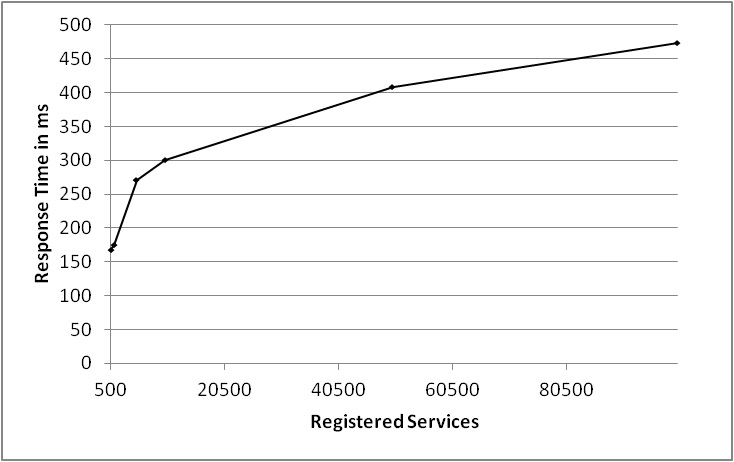} 
\caption{Service Discovery Time for varied registered services}
\label{fig:scale1}
\end{figure}

\begin{figure}[!h]
\centering
\includegraphics[scale=0.5]{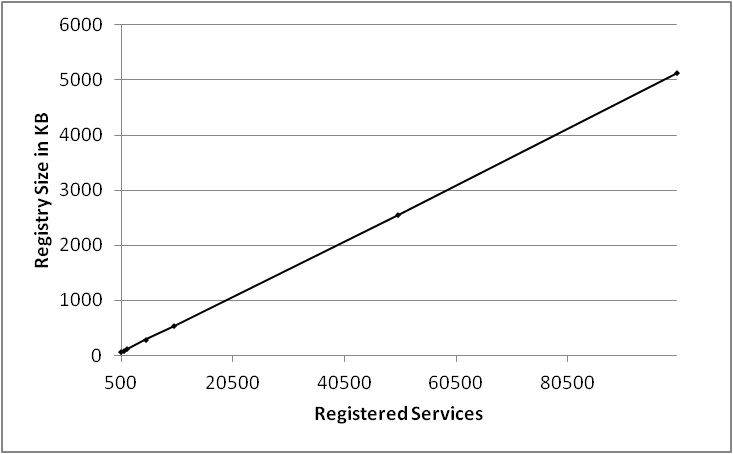} 
\caption{Registry size for varied registered services}
\label{fig:scale2}
\end{figure}

We sent service discovery requests from four mobile devices and virtual instances 
to varied numbers of registered services and the response time was calculated for these requests. Figure~\ref{fig:scale1} shows the average response time for these discovery requests for various registry sizes.
Through the whole experiment, the mobile device acting as the registry node was moving continuously within the network. Further, the size of the registry increased linearly with the increasing number of registered services. Even with thousands of services registered, the registry size was under 10 MB (shown in Figure~\ref{fig:scale2}). It should be noted here that we made use of SQLite for implementing the service and group registries over android mobile devices.

\begin{figure}[!h]
\centering
\includegraphics[scale=0.5]{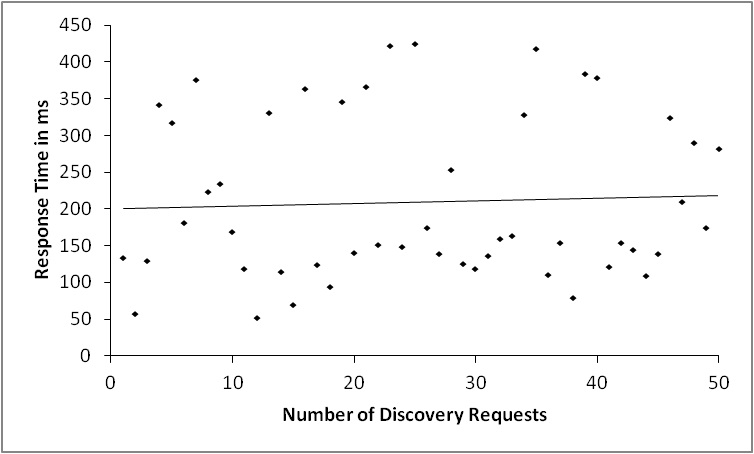} 
\caption{Response Time behavior without an active call}
\label{fig:wocall}
\end{figure}

\begin{figure}[!h]
\centering
\includegraphics[scale=0.5]{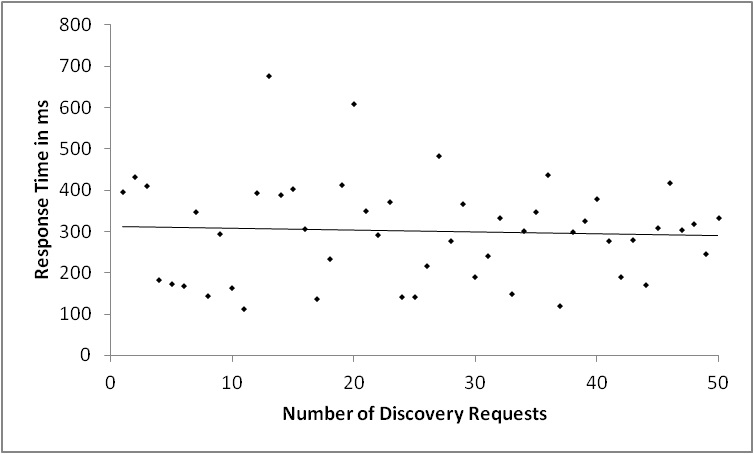} 
\caption{Response Time behavior during an active call}
\label{fig:call}
\end{figure}

Our fifth experiment aimed at evaluating the feasibility of the proposed approach when running simultaneously with other phone activities. In this experiment, we conducted a comparative study on the difference in response time for the service discovery requests a) when a volunteer was answering a phone call and, b) when the device was idle. 
For this experiment, we sent 50 requests for group discovery to the volunteer's mobile device. First, we observed the response behavior when the device was idle in the volunteer's pocket. During this phase the volunteer was randomly moving within the network. 
The response time behavior during this period is shown in Figure~\ref{fig:wocall}.
Next, we made a phone call on the volunteer's mobile phone and observed the response time during the call.
The results demonstrate that there is a change in response time, but this change is well within acceptable limits. Response time behavior during the call is shown in Figure~\ref{fig:call}.
There is an initial peak in the response time when the call is made.
From the initial results, we conclude that the mobile device 
takes a little time to respond to the first discovery request. This is due to the fact that some processing time is required to ~\textit{awaken} the sleeping mobile application. We feel, therefore, that ~\textit{piggybacking of incoming requests} could be a good approach to reduce the energy overhead.


The sixth experiment involved an evaluation of the reliability of the approach.
We toggled the availability status of the service provider from~\textit{available} to~\textit{unavailable} and back to~\textit{available} with a time difference of 10 seconds. These toggles were repeated 120 times at the registry node. 
During this time, we continuously probed the registry node from the service consumers to get the status of the service provider. 
The initial results shown by the experiment have less than 1\% false negative, where false negative implies - registry node returns availability information as~\textit{available} when the service provider has updated it to~\textit{unavailable} and vice versa.

Although the scale of these experiments was limited, the results were promising. 
It will be interesting to explore the performance of this approach for 
a much larger number of service providers, and registry clients and for much longer durations (a few days).
(Interested readers may refer to https://goo.gl/4VG895 for further details about the architecture and experimental setup.)

Nonetheless, we present the speculated trend of our experiments with a large number of devices:
\begin{enumerate}
\item Effects on battery: With increasing number of devices there would be more numbers of registry requests and responses; hence more battery power would be needed. However, in such scenarios the service groups and navigator nodes would play a crucial role. More specialized service groups would be formed as suggested in section~\ref{subsec:concept}, this would keep the number of devices in a service group within limits in turn keeping the battery usage in check (irrespective of any number of registered services).
\item Effects on data exchange: With increasing number of devices there would be more exchange of data. However, the split service groups would keep a check on the data exchange. Further, keeping in view the current data usage by modern smartphones, the data usage by large number of devices should be in acceptable limits.
\item Effects on service registration time: We believe that with increasing number of devices and service registrations, the response time would increase but would be well within the accepted range. Also, this is a onetime process for a service, it would be acceptable for all practical purposes.
\item Effects on service discovery time: Figure~\ref{fig:scale1} shows the discovery time for a large number of registered services varying from one thousand to a few hundreds thousand. This is still within acceptable limits.
\item Effects on registry size: The registry size will increase linearly for increasing number of register services. Trend is shown in the Figure~\ref{fig:scale2}.
While other experiments are dependent on the individual mobile device without depending on the other device’s behavior. Further, the effects of other contextual parameters on increasing number of devices (e.g. network, ISP, terrain, carrier type etc.) would be interesting to look into but unfortunately these are not within the scope of current focus.
\end{enumerate}

\subsection*{Discussion}
\label{subsec:discuss}
The proposed architecture caters to the issues of providing a dynamic service registry in mobile environments. It manage to incorporate the requirements outlined in Section~\ref{sec:req}. 

\textit{R1: Management of transient web services:} The approach effectively manages the availability information on each registered web service and is capable of dynamically updating it. This helps in satisfying requirement~\textit{R1} and keeping the service registry up-to-date irrespective of random entry and exit of transient web services. 

\textit{R2: Lightweight:}  The architecture seeks information that is just enough to uniquely identify and manage a web service from the registrant mobile device for registry related operations. The registries manage information that is less likely to change and whatever change does happen is updatable via a watchdog process. This keeps the registry architecture as lightweight as possible and thus satisfies requirement~\textit{R2}.

\textit{R3: Minimum communication overhead:} We have made use of just three XML stanzas in order to minimize communication overhead, satisfying requirement~\textit{R3}. During our experiments, we observed the exchange of just a few kilobytes of data for hundreds of request transfers. This small overhead also helped us minimize battery utilization, as reflected in Table~\ref{table:power} and~\ref{table:network}.
Furthermore, the approach uses just one stanza ``IQ Stanza'' for easier service registration and de-registration.

\textit{R4: Distributed Service Registry:} The proposed approach manages the service registry over dispersed registry nodes and navigator nodes. 
Thus satisfying requirement~\textit{R4} and improving fault localization. These features, contribute to a light weight and autonomous service registry effective for mobile environments.

\textit{R5: Run time search:} The dynamic availability information, minimal data transfer, and faster response time helped in performing effective run time searches and in the process satisfied requirement~\textit{R5}. 

Further, the proposed dynamic mobile service registry is compatible with existing UDDI and other registries. The dynamic registry can register external UDDI and other service registries as one of the registered services along with their access points. Contrary the proposed registry can be registered with other registries and UDDI as any of the registered service.

\subsection*{Assumptions and Limitations}
\label{subsec:assump}

There are certain important assumptions that have been made in the proposed service registry framework. These are listed as follows:
1) The load balancing of the incoming registry requests is assumed to be handled at the device level. There could be a threshold capacity limit, at each device depending on its hardware capacity. On exceeding this capacity limit, the incoming requests are not entertained and these surplus requests are handled by other registry nodes that have access to the common access channel. 
2) There is no reward system suggested in the current state of the work. Hence, it is assumed that the mobile device owners are self-motivated to provide their respective devices for serving as the registry. We may look into a system of rewards/incentive as part of future work. 
3) The updates are performed to keep the local registry replica updated (at the navigator/registry nodes) in an automated manner without manual intervention.

A few important limitations of the current work are listed as follows:
a) The current design of the framework does not deal with the privacy issues of the mobile phone users associated with service registry and service provisioning. 
b) The current design of the mobile registry does not handle the QoS (Quality of Service) aspect of the mobile web service. The QoS may be handled by providing a link to the data server (external to the framework) that could have QoS and other details on the service description.
c) Though we have a system in place to detect the unavailability of registry nodes through heartbeat signals, the behavior of the mobile device owner, poor network connectivity, and physical damage to the mobile device could result in the abrupt unavailability of the registry/navigator node. These have not been dealt with in this work. 
d) Finally, the current evaluation of the approach was conducted within a supervised lab environment. The registry power and data requirements, service discovery performance, and other results were therefore well within acceptable range. There could be slight variations in these if the experiments were carried out at a much larger scale including thousands of mobile devices.

\section{Related Work}
\label{sec:related}

Service discovery is an important aspect of service-oriented architecture. Two types of approaches are primarily adopted for service discovery: Registry based Approach and Registry-less Approach. The Registry-less approach usually makes use of overlay networks, hash tables, and other broadcasting/multi-casting techniques. 
In the proposed work, we perform service discovery using the former i.e. "Registry based Approach". 

We surveyed existing literature from the late 90's. We classified the registry based approach into two broad categories: the Centralized Registry Approach and the Distributed Registry Approach. These two can further be classified into those for mobile environments and those for non-mobile environment. Our survey includes works from the areas of SOA, peer-to-peer networking, mobile ad-hoc networking.


\subsection{Centralized Service Registry:}

The centralized service registry approach is used in several popular technologies: 
Service Location Protocols~\citep{guttman1999service}, Sun's Jini architecture~\citep{waldo1999jini}, Service Discovery Services~\citep{czerwinski1999architecture}, Microsoft's Universal Plug and Play (UPnP)~\citep{miller2001home}. 
These service discovery infrastructures rely on a central registry for discovering capable services. The service information is stored at a centralized registry. All registry related operations are performed by a single entity.

One well known example of centralized service registry architecture for web services is UDDI (Universal Description Discovery and Integration)~\citep{uddi302}. 
UDDI is not the service registry itself. However, UDDI is the specification of a framework for describing web-services, registering web-service, and discovering web-service. 
Several data structures and APIs have been published for describing, registering and querying web-services through the UDDI.
The ebXML (electronic business XML)~\citep{hofreiter2002ebxml} standard is another example of a centralized web service registry architecture. Hoschek \citep{hoschek2003} presented a grid based hyper registry for web services in peer-to-peer networks. The registry is an XQuery based centralized database that manages dynamic distributed contents.

Juric et al.\citep{Juric2009} proposed an extension to the UDDI for incorporating version support for services. They presented modifications to the category tag of business service and tModel of the UDDI infoset with the intent to introduce service interface versions in UDDI and WSDL.
Bernstein and Vij~\citep{bernstein2010intercloud} proposed the use of XMPP for intercloud topology, security, authentication and service invocations. In some ways this work is similar to ours. However, our work focuses on registry management in mobile based service oriented architecture. We focus
on managing the service registry in a distributed manner over resource constrained mobile devices.
Seto et al.~\citep{seto2011} proposed a service registry for ubiquitous networks to dynamically discover service resources. Their registry divides the service operation into the source, transformation, and sink, specifies physical meta-data to manage devices, and associate a keyword with it.

Feng et al.~\citep{Feng2011} proposed a registry framework to include interoperability among various semantic web service models. For this, they made use of registry meta model for interoperability-7 and several mapping rules for handling semantic mismatch between services.
Feng et al.~\citep{feng2013} proposed a service evolution registry where providers can register their service evolution information and consumers can be sent an alert regarding the service evolution. Their work manages service versions along with their dependencies and discrepancies. 

More recently~\citep{exten2015raey} presents the idea of using object relational databases in the service registry. The registry extends the search to include search on the basis of service commitments and service expectations. 

 \subsection{Centralized Mobile Service Registry:}

Some existing work~\citep{diehl1996}\citep{james1999} discusses the possibility of centralised service registries for mobile environments. Diehl et al.~\citep{diehl1996} talk about centralized service registries that store the service domain, service types, location and access rights to manage service mobility and adoption of services in wireless networks. Beck et al.~\citep{james1999} propose an adaptable service framework for mobile devices that relies on a central service registry for dynamic service registration and discovery. 

Doulkeridis et al. \citep{doulkeridis2003} discuss the idea of managing contextual information in service registries. The main focus of the work is to ease service discovery in mobile environments by maintaining context aware service registries. 
Deepa and Swamynathan~\citep{deepa2010} talk about a directory based architecture that make use of two integrated architectures: backbone-based and cluster based. Their work was intended to facilitate service discovery in mobile ad-hoc networks and to achieve improved network traffic, response time, and hit ratio.


\subsection{Distributed Service Registry:}

Chen et al.~\citep{chen2003} and Sivashanmugam et al.\\~\citep{sivashanmugam2004} discuss a
few initial approaches to maintain UDDI in a federated environment. 
One approach ~\citep{chen2003} supports QoS based discovery from requesters and provides an aggregated result from the federated registries. The other approach~\citep{sivashanmugam2004} suggests the use of various metadata and ontologies to manage UDDI in a federated environment.
Verma et al.\citep{kunal2005} propose METEOR-S WSDI, that focusses on providing registries in distributed and federated environments. Extended Registries 
Ontology was used to provide access to these distributed registries and organizing them in domain based categorization.

The approach discussed in~\citep{Bubak2005} presents a distributed service registry for grid application. The approach utilises Xpath queries and ontological trees for domain based service discovery. 
Baresi and Miraz~\citep{baresi2006} talk about an approach to enable heterogeneous federated registries to exchange service information. The approach is based on the publish and subscribe model.
Ad-UDDI~\citep{du2006} is a distributed registry architecture that adopts an active monitoring 
mechanism. The approach extends UDDI to incorporate automated service updates in a federated registry environment.

Treiber and Dustdar~\citep{treiber2007} propose an active web service registry that make use of atom news formats. RSS software is used in the approach to form an active distributed registry.
Shah et al.~\citep{shah2010} also propose an RSS-based distributed service registry in the move to achieve global SOA. The proposed registry is intended to provide dynamic discovery using RSS and tries to resolve synchronization issues in RSS. 
Jaiswal et al.~\citep{jaiswal2013} introduce a decentralized registry using the Chord protocol for peer-to-peer environments. The registry comprises distributed hash tables of web-service names and web-service IPs. Their method claims to cater to demand driven web-service provisioning.

Another direction in distributed service registry systems is one meant for cloud environments~\citep{lin2013} and \citep{Elgazzar2014}. Lin et al. \citep{lin2013} present a hadoop-based service registry for the cloud environment. The work proposes geographical knowledge service registries that are designed to simplify service registration, improve discovery and other registry operations for cloud services.
Elgazzar et al.~\citep{Elgazzar2014} propose to manage a local service registry at the provider's site for the offered services. This local service registry is managed in a distributed manner and has two types of services: local and remote. The paper proposes Discovery-as-a-Service in the cloud environment.

Das Gupta et al.~\citep{dasGupta2014} in a more recent paper, discuss about the possibility of a federated registry system for P2P networks. The work makes use of multi-agent based distributed service discovery for non-deterministic and dynamic environments. They propose that super peer nodes manage the distributed service registry and other peers register their services with these registries. 
Zhang et al.~\citep{zhang2014} discuss the integration of peer to peer technology with  SOA. They talk about self-organizing, semi-structured P2P frameworks for support and propose to use a private service registry at each peer for discovering the manufacturing services. The local private registry of the peer is traversed first to discover a service.

\subsection{Distributed Mobile Service Registry:}

Handorean and Roman~\citep{handorean2002} propose probably the first work that discusses the possibility of distributed service registries in mobile environments. In their approach, the availability of services is shown in the registry along with an atomic update facility to maintain consistency. 
Konark~\citep{helal2003} is a distributed XML based registry that has a tree structure. A top-down approach is used for the tree, with generic classification of services at the top and specific classification at the bottom. Every node  maintains  a  service  registry, where  it  stores  information  about  its  own  services  and  also  about  services  that  other nodes  provide. The approach provides a semantic service registry and enables servers as well as clients to register and discover services in a distributed manner.

Schmidt and Parashar~\citep{schmidt2004} present a distributed hash table based approach for distributed registries in peer to peer networks. The approach supports service discovery based on keywords, wild cards on an Internet scale. Indexing of keywords associated with the web service description document is managed at the peers.
Tyan and Mahmoud~\citep{tyan2005} discuss an approach for service discovery in mobile ad-hoc environments.  The work considers a registry as a tree and makes this registry available to every node in the network. The approach makes use of a location aware routing protocol and divides the network into hexagonal grids with a gateway for each that have the service registry. 
Golzadeh and Niamanesh~\citep{golzadeh2011} discuss an approach for a service registry system for mobile ad-hoc networks. The approach divides the MANET into clusters with one head for each cluster that acts as a directory for the cluster. The head node has two types of registries: the service provider's registry and the other head node registry.

A decentralized service registry for location based web services is discussed in~\citep{dsouza2014}. The approach is relied on cellular network system and base transceiver station for retrieving the local registry address. These addresses are broadcasted by base station and interested mobile devices download the registry address for location based services.
One of the latest works by Jo et al.~\citep{distri2015jo} makes use of bloom filter to manage distributed service registry for mobile IoT environments. Proposed work uses hierarchical bloom filters for reducing message exchanges among registries in move to find available services. 

Although there is a good amount of work that has been done towards developing effective service registries, 
an architecture that enables mobile devices to host mobile service registries and that
takes the distinct features of mobile environment into account such as intermittent connectivity, dynamic nature, frequent service description changes is still lacking. 
Our work focuses on registry management in mobile based service oriented architecture. We focus on managing the service registry in a distributed manner on resource constrained mobile devices.
This service registry contains minimal information about the registered services in a manner that is just enough to uniquely discover the services. The proposed work is an extension of some of our previous work~\citep{rohit2014ws}.

\section{Conclusions}
\label{sec:conclude}

In this paper we looked into, perhaps, the most challenging aspect of implementing an SOA in mobile environments: effective service registries. 
Our studies show that traditional approaches for implementing service registries (such as UDDI) cannot be directly adopted in mobile environments, given the dynamic, volatile and uncertain nature of such environments.
A novel approach to manage service registries `solely' over mobile devices was proposed that effectively addressess issues specific to mobile environments and enables run time service discoveries.

We evaluated the approach by developing a prototype and deploying it over real mobile devices.
To emulate real world usage as closely as possible we requested volunteers to deploy our prototype on their personal mobile devices and continue doing their routine tasks. 
The experimental results indicate that the proposed solution is an effective enabler for SOA in mobile environments.
We performed several experiments to confirm the efficacy of the prototype across several parameters such as: timely performance, battery consumption, effect of the random/nomadic behaviour of people carrying mobile devices, conflict with native mobile apps, reliability.

Future work in this direction would be towards mobile service registries that focus on QoS factors unique to mobile environments. Future work will also tackle security related issues, dynamic service group splitting in mobile service registries. Further, the availability of the service registries will be a prime focus.

\section*{Acknowledgment}
We would like to thank Tanveer Ahmed and Dheeraj Rane for their valuable insights.
The work was supported by the Ministry of Human Resource Development - Government of India. 


\section*{References}

\end{document}